\documentclass[referee]{jfm}
\usepackage[dvips]{graphicx}
\usepackage{amsmath,amssymb,amsbsy,epsfig}
\usepackage[a4paper, left=2cm, right=2cm, top=2cm, bottom=2cm]{geometry}
\usepackage{natbib}

\title[Wetting front in a porous medium]
{Wetting front dynamics in an isotropic porous medium}

\author[Y.D. Shikhmurzaev and J.E. Sprittles]
{Y\ls U\ls L\ls I\ls I\ns D.\ns S\ls H\ls I\ls K\ls H\ls M\ls U\ls
R\ls Z\ls A\ls E\ls V\footnote{E-mail: Y.D.Shikhmurzaev@bham.ac.uk}
\and J\ls A\ls M\ls E\ls S\ns E.\ns S\ls P\ls R\ls I\ls
T\ls T\ls L\ls E\ls S\footnote{E-mail: sprittlj@maths.bham.ac.uk}}

\affiliation{School of Mathematics, University of Birmingham, Birmingham,  B15 2TT, UK.}

\begin{document}

\label{firstpage} \maketitle

\begin{abstract}
A new approach to the modelling of wetting fronts in porous media
on the Darcy scale is developed, based on considering the types
(modes) of motion the menisci go through on the pore scale. This
approach is illustrated using a simple model case of imbibition of
a viscous incompressible liquid into an isotropic porous matrix
with two modes of motion for the menisci, the wetting mode and the
threshold mode. The latter makes it necessary to introduce an
essentially new technique of conjugate problems that allows one to
link threshold phenomena on the pore scale with the motion on the
Darcy scale. The developed approach (a) makes room for
incorporating the actual physics of wetting on the pore scale, (b)
brings in the physics associated with pore-scale thresholds, which
determine when sections of the wetting front will be brought to a
halt (pinned), and, importantly, (c) provides a regular framework
for constructing models of increasing complexity.
\end{abstract}

\section{Introduction}

The dynamics of wetting fronts in porous media remains the subject
of intensive research
\citep{Brenner-review88,Olbricht-review-1996,Alava-review04}. Its
main motivation comes, first of all, from a host of important
applications, notably in oil recovery, hydrogeology and more
recently also in carbon dioxide sequestration, microfluidics and
fuel cells. This topic also poses some fundamental questions about
the modelling of evolutionary processes in systems with complex
topology. In a practically relevant case where the scales of the
pore-level and the global flow are well separated, one can use an
intermediate scale to introduce averaged macroscopic quantities and
apply the modelling approach of continuum mechanics. In this case,
for an incompressible viscous liquid invading an isotropic porous
medium the averaged flow velocity $\mathbf{u}$ and pressure $p$,
both functions of the position vector $\mathbf{r}$ and time $t$,
satisfy the equation of motion in the form of Darcy's law
\begin{equation}
\label{Darcy-1}
 \mathbf{u}=-(\kappa/\mu)\nabla p,
 \qquad(\mathbf{r}\in\Omega),
\end{equation}
where $\kappa$ and $\mu$ are the permeability of the porous matrix
and the fluid's viscosity, respectively, and $\Omega$ is part of
the porous medium occupied by the fluid. Darcy's law together with
the continuity equation $\nabla\cdot\mathbf{u}=0$ form a closed
system adequately describing, on the macroscopic level (`Darcy
scale'), the bulk distributions of $p$ and $\mathbf{u}$. Combining
these two equations, one has that the pressure in the flow domain
$\Omega$ is harmonic
\begin{equation}
\label{Laplace-p} \nabla^2p=0,\qquad(\mathbf{r}\in\Omega),
\end{equation}
so that, to determine $p$, one has to specify two boundary
conditions on the part of the boundary of $\Omega$ whose location
is unknown (i.e.\ the wetting front; hereafter $\partial\Omega_1$)
and one boundary condition on the part whose position is known
(hereafter $\partial\Omega_2$, so that
$\partial\Omega_1\cup\partial\Omega_2=\partial\Omega$ is the
boundary of $\Omega$). Hence on the wetting front, in addition to
the kinematic condition
\begin{equation}
\label{kinematics-general}
 \frac{\partial f}{\partial t} + \mathbf{u}\cdot\nabla f=0,
 \qquad(\mathbf{r}\in\partial\Omega_1),
\end{equation}
which specifies the evolution of $\partial\Omega_1$ in terms of
its {\it a priori\/} unknown location $f(\mathbf{r},t)=0$, we need
to formulate an appropriate dynamic boundary condition for $p$. If
the dynamics of the displaced fluid also needs to be considered,
as, for example, in the case of one viscous liquid displacing
another, one will still need a dynamic boundary condition relating
the two fluids' pressures at the interface.

The main issue in determining the dynamic boundary condition is to
what extent the actual physics of wetting on the pore scale is
accounted for, and how it is represented, on the Darcy-scale
level. In particular, as has been known for a long time
\citep{Huh-Scriven71,Dussan-Davis74,Dussan-review79}, the
classical model of fluid mechanics does not allow viscous fluids
to spread over a solid surface with a contact angle less than
$180^\circ$, whereas numerous experiments show that they do
\citep[e.g.\ see Ch.~3 of][]{TheBook}. The way one chooses to
overcome this (`moving contact-line') problem for the menisci that
collectively form the wetting front in a porous medium is one of
the factors determining how realistic the resulting model will be.
At present, this aspect of the wetting front dynamics modelling
received almost no attention, even in the approach which considers
the porous medium as a network of capillaries and hence is
potentially capable of capturing exactly the details of the
pore-scale physics \citep{Lenormand88,Aker00,Joekar-Niasar-2010}.

An alternative to the topologically transparent sharp-interface
approach outlined above is to consider the wetting front as a
transition zone where the volumetric concentration of the
displaced and displacing fluid change, with no distinction between
continuous and discrete phases, and treat this, essentially
multiphase, system as a multicomponent one
\citep{Richards-1931,Leverett-1941}, using a thermodynamic closure
to relate the pressure difference between the two fluid phases
with saturations and, possibly, other variables
\citep{Hassanizadeh-Gray-1993,Mitkov-Tartakovsky-1998,Deinert-etal-2008}.
A difficulty in this approach is that, as experiments aimed at
determining the closing thermodynamic relationships indicate the
need to bring in more and more parameters of state, it is not
clear whether or not thermodynamics is an adequate tool to
describe this, essentially mechanical, system.

\section{\label{TheModel}The model}

In this paper, we introduce a new approach to formulating models
describing the propagation of liquid-fluid interfaces across porous
media based on considering the {\it types (modes) of motion\/} which
the menisci that collectively form the free surface go through as
they advance across the porous matrix. This approach offers a
regular way of building models of increasing complexity accounting
for the topological and geometric features of the porous medium. It
is worth noting that Darcy's equation (\ref{Darcy-1}) in the bulk is
itself essentially a consequence of the flow profile  being
approximately parabolic on the pore scale, i.e.\ it also can be seen
as based on a particular type of flow. In this sense, the approach
we are developing here considers the boundary conditions
conceptually in the same way as the bulk equations.

\begin{figure}
\centerline{\epsfig{file=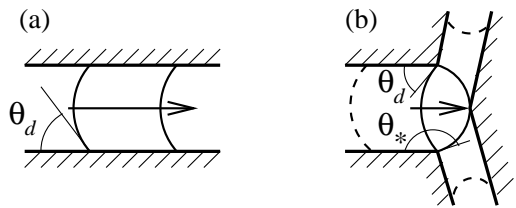,scale=0.45}}
 \caption{Schematic illustration of the meniscus motion in the wetting mode (a) and
  one of scenarios associated with the threshold mode (b).}
\label{sketch}
\end{figure}

We will illustrate the new approach using the simple case of a
viscous incompressible liquid displacing an inviscid dynamically
passive gas from an isotropic homogeneous porous medium.  On the
pore scale, each meniscus intersects the pore boundary at a
`contact line', forming a certain `contact angle' with the solid.
We can schematically represent the motion of the meniscus on the
pore scale as having two principal modes: (i) the {\it wetting
mode}, where the contact line moves across the pore boundary with
negligible variation in the meniscus shape, and (ii) the {\it
threshold mode}, where the contact line becomes pinned whereas the
meniscus deforms until the contact angle it forms with the solid
reaches a critical value at which the contact line can move again
(Fig.~\ref{sketch}). These two modes control the motion as each of
them is capable of bringing individual menisci and hence the
wetting front as a whole to a halt. In the wetting mode this can
happen when the contact angle becomes equal to the equilibrium
one, so that the meniscus no longer needs to move to reach an
equilibrium state, whereas in the threshold mode, where the
contact line is pinned, the flow stops when the pressure building
on the meniscus (later referred to as
$\bar{p}|_{\partial\Omega_1}$) is insufficient to break through
the threshold.

Macroscopically, for the pressure on the wetting front measured
with respect to the (presumed constant) pressure in the displaced
gas one has
\begin{equation}
\label{p=} p|_{\partial\Omega_1}=A_1 p_1+A_2 p_2,
\end{equation}
where $p_1$, $p_2$ are the averaged pressures and $A_1$, $A_2$ are
the spatio-temporally averaged fractions of the unit area of the
free surface corresponding to the two types of motion
($A_1+A_2=1$). Importantly, $A_1$ and $A_2$ do not have to remain
constant as the wetting front propagates, and they are yet to be
specified.

We will begin by considering the wetting mode. For simplicity, we
will represent the pore where the wetting motion takes place as
having a circular cross-section. Then, for low capillary and
pore-scale Bond numbers, the meniscus will have the shape of a
spherical cap and
\begin{equation}
\label{p_1=} p_1=-2\sigma\cos\theta_d/a,
\end{equation}
where $\sigma$ is the liquid-gas surface tension, $a$ is the
effective radius of the pore and $\theta_d$ is the {\it dynamic\/}
contact angle that the meniscus forms with the solid wall. As
shown by experiments on dynamic wetting, in a general case
$\theta_d$ depends on, and hence should be regarded as a {\it
functional\/} of, the flow field in the vicinity of the moving
contact line \citep{Blake-Bracke-me99,Clarke-assist06}. For flow
in a porous medium, the contact angle's dependence on the flow
field reduces to its dependence only on the contact-line speed
$u_1$, which is the leading factor determining the flow field in
the vicinity of the contact line. Then, $\theta_d$ can be
described just as a {\it function\/} of the contact-line speed:
\begin{equation}
\label{theta-empirical}
 \theta_d=F(u_1/U_{cl}),
\end{equation}
where $U_{cl}$ is an appropriate scale for the velocity. In
principle, this dependence, where, for the wetting mode, $u_1$
coincides with the speed of the meniscus as a whole, could be
determined empirically. The theory of flows with forming
interfaces \citep{TheBook} specifies the inverse of $F(u_1/U)$ as
\begin{equation}
\label{theta_d=}
\frac{u_1}{U_{cl}}=
 \left(
 \frac{
 (1+(1-\rho^s_{1e})\cos\theta_s)(\cos\theta_s-\cos\theta_d)^2}
 {4(\cos\theta_s+B)(\cos\theta_d+B)}
 \right)^{1/2},
\end{equation}
where $B=(1-\rho^s_{1e})^{-1}(1+\rho^s_{1e}u_0(\theta_d))$,
$\theta_s$ is the static contact angle,
$$
 u_0(\theta_d)=\frac{\sin\theta_d-\theta_d\cos\theta_d}
 {\sin\theta_d\cos\theta_d-\theta_d},
 \qquad
 U_{cl}
 =\left(\frac{\gamma\rho^s_0(1+4\alpha\beta)}{\tau\beta}\right)^{1/2}
$$
is the characteristic speed associated with the parameters that
the `additional' physics of wetting brings in to resolve the
moving contact-line problem, $\rho^s_0$, $\rho^s_{1e}$, $\alpha$,
$\beta$, $\gamma$, $\tau$ are material constants characterizing
the contacting media whose values can be found elsewhere
\citep{Blake-me02,TheBook}. The comparison of (\ref{theta_d=})
with experimental data published in the literature has shown a
very good agreement \citep{TheBook} so that (\ref{theta_d=}) can
be regarded as a reliable representation of the dynamic contact
angle behaviour.

An implicit assumption we made above, namely that in the wetting
mode the contact-line speed $u_1$ equals to the cross-sectionally
averaged flow velocity $u_{1,flow}$ associated with the motion of
the meniscus as a whole, is not immediately obvious, especially
given that, as the meniscus breaks through the threshold and goes
into the wetting mode, initially, one has both the moving contact
line and the varying shape of the meniscus. In other words, in
general one has
$$
u_{1,flow}=u_1+\frac{a}{(1+\sin\theta_d)^2} \frac{d\theta_d}{dt},
$$
where, as before, we used that the meniscus has the shape of a
spherical cap, albeit with a time-dependent radius of curvature.
Formally, one can add this equation to the model together with an
extra variable $u_{1,flow}$, which should replace $u_1$ in the
text below, but this generalization would be beyond the accuracy
of the model. Indeed, if $L$, $U$ and $T=L/U$ are the length,
velocity and time scales for the macroscopic (Darcy-level) motion
for which we are deriving the boundary conditions, then one has
that the difference between $u_{1,flow}$ and $u_1$ is of $O(a/L)$
and hence negligible in the continuum limit $a/L\to0$. The
possible deviation of the meniscus shape from a spherical cap
takes place also on a vanishing scale. Therefore, within the
accuracy of $O(1)$ as $a/L\to0$, in what follows we use that in
the wetting mode $u_{1,flow}=u_1$. It is worth pointing out here
that the pores (i.e.\ capillaries) where the meniscus propagates
in the wetting mode are assumed to be long compared to $a$, so
that there is room for the meniscus to propagate in the wetting
mode as it is described above.

The speed $u_1$ at which individual menisci propagate in the
wetting mode is not equal to the normal component of the velocity
of the wetting front as a whole
$u_n=\mathbf{n}\cdot\mathbf{u}|_{\partial\Omega_1}$ ($\mathbf{n}$
is an outward normal) since the menisci intermittently go through
both modes of motion and hence, on the Darcy scale, $u_n$ must
have contributions from both $u_1$ and the flow speed $u_2$
associated with the threshold mode. Then, for $u_n$ one has an
equation
\begin{equation}
\label{u_n=} u_n=A_1u_1+A_2u_2,
\end{equation}
which is similar to (\ref{p=}), with the contribution of the {\it
i\/}th mode proportional to its `weight' $A_i$. In an isotropic
porous medium, the `weight' $A_i$ of each mode of motion is
essentially the relative time the meniscus spends in this mode. If
$s_i$ is the fraction of the length on the pore scale
corresponding to the $i$th mode of motion $(s_1+s_2=1)$, then the
normalized time that the meniscus spends in this mode is $s_i/u_i$
and hence, given that $1/u_n=s_1/u_1+s_2/u_2$ and $A_1+A_2=1$, one
has
\begin{equation}
\label{alphas}
A_1=\frac{s_1u_2}{s_2u_1+s_1u_2},
\quad
A_2=\frac{s_2u_1}{s_2u_1+s_1u_2}.
\end{equation}
Then, as one would expect, the slowest (controlling) mode of
motion tends to make a greater contribution to the pressure at the
wetting front and the front's velocity.

Now, consider the threshold mode. When the moving meniscus runs
into a barrier associated with the threshold mode, such as the
edge at the end of a capillary or an asperity, the contact line
gets pinned and the meniscus begins to deform until the contact
angle reaches a certain value $\theta_*$ at which the contact line
can move forward again (Fig.~\ref{sketch}). This is the essence of
the threshold mode, and it comes into play only when $\theta_d$ at
which the meniscus arrives at a barrier is less than $\theta_*$.
In other words,
\begin{equation}
\label{s=}
 s_1(\theta_d,\theta_*)=
 \left\{
 \begin{array}{ll}
 1, & \theta_d-\theta_*\ge0\\
 s_{10}, & \theta_d-\theta_*<0
 \end{array}
 \right.,
 \qquad
 s_2=1-s_1,
\end{equation}
where $s_{10}$ $(<1)$ is a characteristic of the porous matrix. To
find the functional dependence of parameters in the threshold
mode, consider the dynamics of an individual meniscus. Assume that
at a distance $l$ upstream from the meniscus that has just run
into a barrier and got its contact line pinned there is a
(constant throughout the process) pressure $\bar{p}$ greater than
$-2\sigma\cos\theta_d/a$. Then, the meniscus, with the contact
line which is now unable to move, will give in to this pressure
and deform (Fig.~\ref{sketch}). We need to find the flow velocity
$u_2$ as an average over the cross-section and over the time
required for the the contact angle to vary from $\theta_d$, at
which the meniscus arrived at the barrier, to $\theta_*$, at which
the contact line can advance again, and the pressure $p_2$ as an
average over the time of this process.

Neglecting the contribution to the pressure drop due to the
deviation of the velocity profile on the pore scale from parabolic
in the immediate neighboorhood of the meniscus and calculating the
flow rate in the capillary on an assumption that the meniscus
retains the shape of a spherical cap (with a varying radius of
curvature) throughout the process of its deformation, from the
Stokes equation on the pore scale one has that the contact angle
$\theta$ satisfies an equation
$$
\frac{1}{(1+\sin\theta)^2}\frac{d\theta}{dt} =\frac{\sigma}{4\mu l}
\left(\frac{\bar{p}a}{2\sigma}+\cos\theta\right),
$$
which, if multiplied by $a$, essentially equates the flow velocity
averaged over a cross-section to the pressure gradient with a
coefficient of proportionality corresponding to the parabolic
profile in the pipe flow. Given that there is no discontinuity in
the average flow velocity when the contact line becomes instantly
pinned and the meniscus starts going from the wetting into the
threshold mode, we have an equation
$$
\frac{\sigma a}{4\mu l}\left(
\frac{\bar{p}a}{2\sigma}+\cos\theta_d\right)=u_1,
$$
which can be used to eliminate $l$. Now, for $p_2$, i.e.\ the
pressure $-2\sigma\cos\theta(t)/a$ averaged over a time interval
$$
T=a u_1^{-1}\left(\frac{\bar{p}a}{2\sigma}
+\cos\theta_d\right) I\left(\theta_d,\theta_*;\frac{\bar{p}a}{2\sigma}\right),
$$
$$
I\left(\theta_d,\theta_*;\frac{\bar{p}a}{2\sigma}\right)
=\int_{\theta_d}^{\theta_*} \frac{d\theta}
{(1+\sin\theta)^2(\bar{p}a/(2\sigma)+\cos\theta)},
$$
which is needed for $\theta$ to vary from $\theta_d$ to
$\theta_*$, one has
\begin{equation}
\label{p_2=}
p_2=\bar{p}-\frac{2\sigma
J(\theta_d,\theta_*)}{aI(\theta_d,\theta_*;\bar{p}a/(2\sigma))},
\end{equation}
where
$$
J(\theta_d,\theta_*)=
\left[\frac{1}{2}\tan\left(\frac{\theta}{2}-\frac{\pi}{4}\right)
+\frac{1}{6}\tan^3\left(\frac{\theta}{2}-\frac{\pi}{4}\right)
\right]_{\theta_d}^{\theta_*},
$$
and $[f]_a^b\equiv f(b)-f(a)$. A similar procedure yields the
velocity in the threshold mode averaged over $T$ as
\begin{equation}
\label{u_2=} u_2=\frac{u_1 J(\theta_d,\theta_*)} {\displaystyle
\left(\frac{\bar{p}a}{2\sigma}+\cos\theta_d\right)
I\left(\theta_d,\theta_*;\frac{\bar{p}a}{2\sigma}\right)}.
\end{equation}
Now, the quantity $\bar{p}$ in equations (\ref{p_2=}) and
(\ref{u_2=}) must be specified in macroscopic terms. On the pore
scale, it is the excess of $\bar{p}$ over the threshold capillary
pressure $p_*=-2\sigma\cos\theta_*/a$ that allows the meniscus to
break through the barrier and go into the wetting regime again,
with the contact line moving. If $\bar{p}$ reduces to $p_*$, then
the meniscus comes to a halt, and, given that all menisci are
modelled as the same, each of them will meet a similar barrier
within a time negligible on the macroscopic scale and the wetting
front as a whole will come to a stop. Formally, we have that, if
$\bar{p}\searrow p_*$, then $I\to\infty$ and hence, according to
(\ref{p_2=}) and (\ref{u_2=}), $p_2\to\bar{p}$ and $u_2\to0$.
Then, we have from (\ref{alphas}) that $A_1\to0$, $A_2\to1$, so
that (\ref{u_n=}) and (\ref{p=}) yield that on the Darcy scale
$u_n\to0$ and $p\to\bar{p}$. Thus, macroscopically (i.e.\ on the
Darcy scale) $\bar{p}$ is the {\it stagnation pressure}, i.e.\ the
pressure that one would have if the wetting front were {\it at
rest in its current position}. In other words, macroscopically, we
need to solve a {\it conjugate problem}
\begin{equation}
\label{conjugate} \nabla^2\bar{p}=0, \quad (\mathbf{r}\in\Omega);
 \qquad
 \mathbf{n}\cdot\nabla\bar{p}|_{\partial\Omega_1}=0,
\end{equation}
with the boundary condition for $\bar{p}$ on $\partial\Omega_2$
being the same as for $p$. Then, the value
$\bar{p}|_{\partial\Omega_1}$ is the one we need to use in
(\ref{p_2=}) and (\ref{u_2=}), i.e.\ it is the pressure that
builds on a meniscus whose contact line has been pinned. The
conjugate problem must be solved in parallel with the main one as
the latter requires the value of $\bar{p}|_{\partial\Omega_1}$ at
the wetting front throughout its movement.

Now, we can summarize the model as follows. In order to describe
the propagation of the wetting front, one has to consider the bulk
equations (\ref{Darcy-1}) and (\ref{Laplace-p}) in the domain
$\Omega$ with some boundary condition on $\partial\Omega_2$ that
specifies a particular problem, with the kinematic boundary
condition at the wetting front $\partial\Omega_1$ given by
(\ref{kinematics-general}) and the dynamic one by (\ref{p=}). The
pressures $p_1$ and $p_2$ that feature in (\ref{p=}) are
determined from (\ref{p_1=}), (\ref{p_2=}), with the coefficients
$A_1$, $A_2$ specified by (\ref{alphas}), (\ref{s=}). For the
three variables $\theta_d$, $u_1$ and $u_2$ appearing in
(\ref{p_1=}), (\ref{alphas}), (\ref{s=}) and (\ref{p_2=}) one has
three equations: (\ref{theta-empirical}) (in particular
(\ref{theta_d=})), (\ref{u_n=}) and (\ref{u_2=}), whereas the
pressure $\bar{p}|_{\partial\Omega_1}$ featuring in (\ref{p_2=})
and (\ref{u_2=}) has to be found from the conjugate problem
(\ref{conjugate}) with the same boundary condition for $\bar{p}$
on $\partial\Omega_2$ as for $p$. If gravity is to be taken into
account, one has to replace $p$ and $\bar{p}$ in (\ref{Darcy-1})
and (\ref{conjugate}) with $(p+\rho gz)$ and $(\bar{p}+\rho gz)$,
respectively ($\rho$ is the fluid's density, $g$ is the
gravitational acceleration, $z$ is the coordinate directed against
gravity). Besides the bulk permeability $\kappa$, the geometry of
the porous matrix enters the model via three effective parameters,
$s_{10}$, $a$ and $\theta_*$. The above model can be generalized
by incorporating threshold modes associated with different values
of $\theta_*$, replacing the step-function (\ref{s=}) for each of
them by a more complex one to account for the types of thresholds
and the corresponding generalization of (\ref{alphas}) to
incorporate various possibilities for the pore network topology.

\section{An illustrative example}

\begin{figure}
\centerline{\epsfig{file=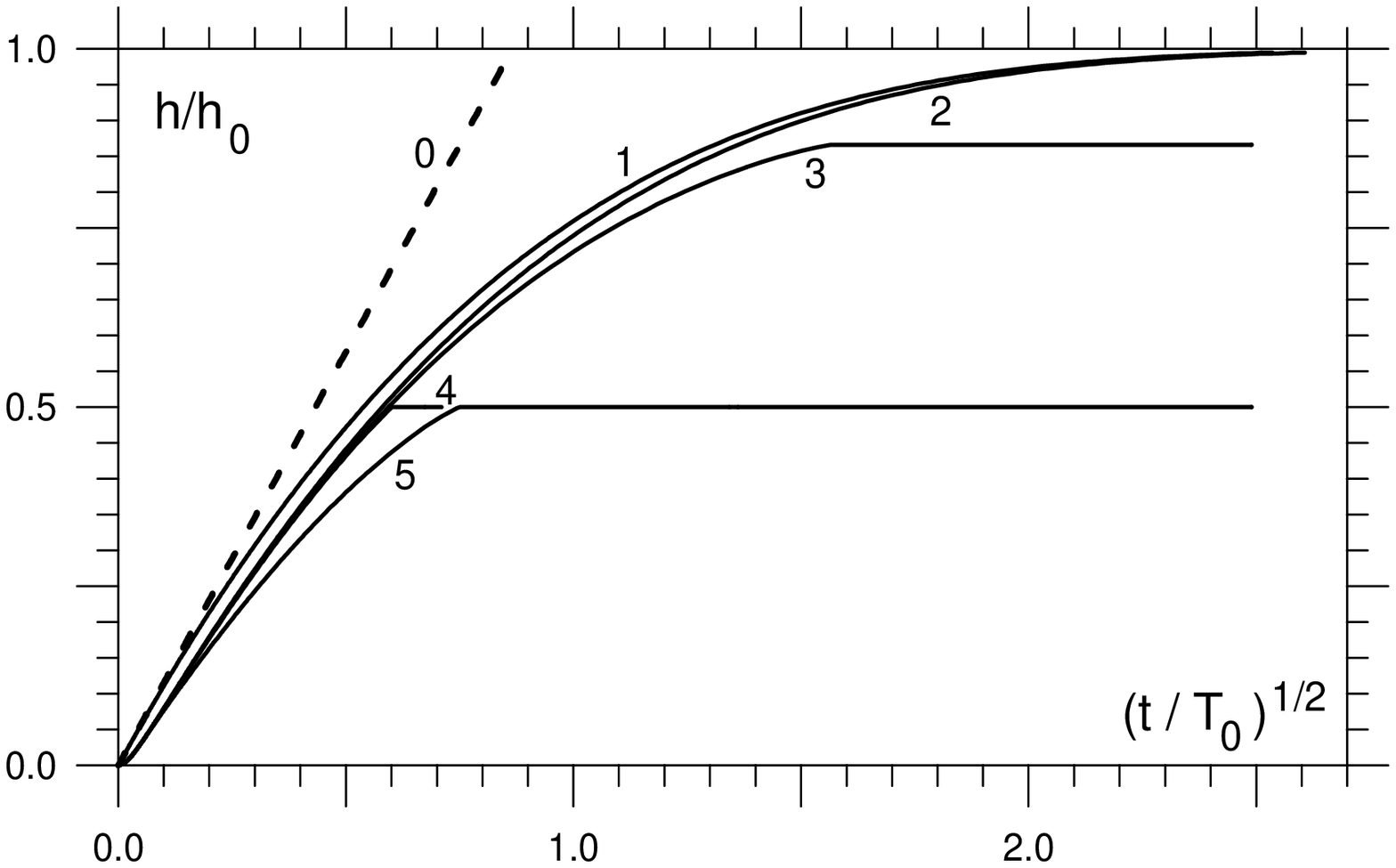,scale=0.5}}
 \caption{Time-dependence of 1D imbibition calculated using the
 derived model.
  0: Washburn, no gravity;
  1: Washburn, gravity included;
  2: $s_{10}=1$;
  3: $s_{10}=0.1$, $\theta_*=30^\circ$;
  4: $s_{10}=0.9$, $\theta_*=60^\circ$;
  5: $s_{10}=0.1$, $\theta_*=60^\circ$.
  For all curves $\theta_s=0^\circ$,
  $p_0=0$, and $\mu U_{cl}/(\kappa\rho g)=10^2$,
  $\rho^s_{1e}=0.6$ for curves 2--5.
   }
\label{rise1d}
\end{figure}

Consider the unsteady one-dimensional imbibition against gravity,
with $z=h(t)$ as the position of the wetting front
($\partial\Omega_1$), $p$ and $\bar{p}$ in (\ref{Darcy-1}) and
(\ref{conjugate}) replaced with $p+\rho gz$ and $\bar{p}+\rho gz$
respectively, and $p=\bar{p}=p_0$ at $z=0$ as a boundary condition
on $\partial\Omega_2$. Then, Laplace's equations (\ref{Laplace-p})
and (\ref{conjugate}) for $p$ and $\bar{p}$ in the one-dimensional
case yield that both $p$ and $\bar{p}$ are linear functions of $z$,
and from the conjugate problem (\ref{conjugate}) with the condition
$\bar{p}=p_0$ at $z=0$ we have that $\bar{p}(z,t)=p_0-\rho gz$.
Then, using for $p$ its linear dependence on $z$ and the same
condition on $\partial\Omega_2$, i.e.\ $p=p_0$ at $z=0$, we can
express the pressure gradient $dp/dz$ at $z=h$ in terms of the
current position of the wetting front: $dp(h,t)/dz=(p(h,t)-p_0)/h$.
Using this expression in Darcy's equation (\ref{Darcy-1}), where, as
mentioned above, $p$ is replaced with $p+\rho gz$, and substituting
the latter into the kinematic condition (\ref{kinematics-general}),
which now can be written down simply as $dh/dt=u_n(h,t)$, we arrive
at
$$
\frac{dh}{dt}=\frac{\kappa}{\mu}\left(
 \frac{p_0-p(h,t)}{h} - \rho g\right).
$$
This equation together with algebraic equations (\ref{p=}), where
$p|_{\partial\Omega_1}\equiv p(h,t)$, (\ref{p_1=}),
(\ref{theta_d=}), (\ref{u_n=}), where $u_n=dh/dt$, (\ref{alphas}),
(\ref{s=}), (\ref{p_2=}), where we now use the solution of the
conjugate problem $\bar{p}|_{\partial\Omega_1}=p_0-\rho gh$, and
(\ref{u_2=}) form a closed system for $h$, $p(h,t)$, $p_1$,
$\theta_d$, $u_1$, $A_1$, $A_2$, $s_1$, $s_2$, $p_2$ and $u_2$.
Typical curves representing the dependence of $h$, scaled with
$h_0=2\sigma/(\rho ga)$ and rising from the initial position
$h(0)=0$, on time $t$, scaled with $T_0=2\sigma\mu/((\rho g)^2
a\kappa)$, are shown in Fig.~\ref{rise1d} with the values of
parameters given in the figure caption. Comparing these dependencies
(curves 2--5) with Washburn's \citep{Washburn-1921} curves for a
meniscus propagating with $\theta_d\equiv\theta_s$ in a capillary
(curve 0: no gravity; curve 1: gravity included), we can see that
the present model, besides realistically giving a finite speed of
rise at the onset of imbibition, predicts a slightly lower rate of
propagation of the wetting front as it is slowed down by both the
velocity-dependence of the contact angle and the presence of the
threshold mode. The latter comes into play once $\theta_d<\theta_*$
and dictates that the wetting front will come to a halt before it
reaches the maximum possible height of imbibition
$h_{max}=2\sigma\cos\theta_s/(\rho ga)$ and, importantly, contrary
to how curves 1 and 2 approach $h_{max}$ asymptotically, this coming
to a halt occurs in a {\it finite\/} time. In this position, which
is determined only by $\theta_*$ (see curves 4 and 5), the wetting
front still has the capacity to propagate provided that some other
physical mechanism helps it to break through the threshold. In this
connection, it is worth pointing out that, as has been observed
experimentally, for some systems, there is a change of regime from
an essentially Washburn-type to a completely different one halfway
between the onset of the process and the maximum imbibition height
\citep{Delker-etal-1996,Lago-Araujo-2001}, so that the present
example, considered here as an illustration of how the developed
approach works, provides a `building block' for the modelling of
this, yet unexplained, phenomenon.

\section{Concluding remarks}

The developed approach and, in particular, the technique of
conjugate problems it uses to incorporate the threshold mode of
motion provide a transparent framework for building models of
increased complexity. The threshold mode is the key to describing
such effects as formation of trapped pockets of the displaced
fluid that can be left behind the wetting front and their
subsequent dynamics without resorting to thermodynamic arguments
and the necessity to specify, increasingly multi-parametric,
thermodynamic dependencies for this, basically mechanical, system.
It should be noted, however, that the next step in the developing
of the model towards greater complexity, sketched at the end of
Section~\ref{TheModel}, is nontrivial as, for different values of
the threshold angle $\theta_*$, it becomes necessary to bring in
and specify the topology of the porous matrix with respect to the
connectivity of the wetting front.

\bibliographystyle{jfm}
\bibliography{references}

\end{document}